\documentclass[conference,letter]{IEEEtran}
\IEEEoverridecommandlockouts
\usepackage{cite}
\usepackage{amsmath,amssymb,amsfonts}
\usepackage{algorithmic}
\usepackage{graphicx}
\usepackage{textcomp}
\usepackage{xcolor}
\def\BibTeX{{\rm B\kern-.05em{\sc i\kern-.025em b}\kern-.08em
    T\kern-.1667em\lower.7ex\hbox{E}\kern-.125emX}}
\usepackage[utf8]{inputenc}
\usepackage{graphicx}
\usepackage{amsmath}
\usepackage{graphicx}
\usepackage{fixltx2e}
\usepackage{mathtools}
\usepackage{listings}
\usepackage{pgfplots}
\usepackage[nolist,nohyperlinks]{acronym}
    
\begin{document}


\title{Adaptive Edge Content Delivery Networks for Web-Scale File Systems}


\author{\IEEEauthorblockN{João Tiago}
\IEEEauthorblockA{
\textit{INESC-ID / Técnico Lisboa - ULisboa, Portugal}
}
\and
\IEEEauthorblockN{David Dias}
\IEEEauthorblockA{
\textit{Protocol Labs}}
\and
\IEEEauthorblockN{Luís Veiga}
\IEEEauthorblockA{
\textit{INESC-ID / Técnico Lisboa - ULisboa, Portugal}
}
}

\maketitle

\begin{abstract}

The InterPlanetary File System (IPFS) is an hypermedia distribution protocol,
    addressed by content and identities. It aims to make the web faster, safer,
    and more open. The JavaScript implementation of IPFS runs on the browser, 
    benefiting from the mass adoption potential that it yields. 
    Startrail takes advantage of the IPFS ecosystem and 
    strives to further evolve it, making it more scalable and performant through 
    the implementation of an adaptive network caching mechanism. 
    Our solution aims to add resilience to IPFS and improve its overall scalability, by avoiding overloading the nodes providing highly popular content, 
    particularly during flash-crowd-like conditions where popularity and demand grow suddenly.
    We add a novel crucial key component to enable an IPFS-based 
    decentralized Content Distribution Network (CDN). Following a peer-to-peer architecture, it runs  on a scalable,
    highly available network of untrusted nodes
    that distribute immutable and authenticated objects which are cached progressively
    towards the sources of requests.
\end{abstract}


\section{Introduction}

In the early days, the Internet was basically a mesh of machines whose main purpose was to share academic and research documents.
It was predominantly a peer-to-peer system, computers connected to the network played an equal role, each capable of contributing with as much as they utilised. It was only due to network topology constraints, mainly NATs, that users on the World Wide Web lost the ability to directly dial other peers. Struggling to overcome obstacles in interoperability of protocols, the gap between client and server nodes widen and the pattern remained. In this pattern, computers play either the role of a consumer - ``client''- or producer - ``server'' - serving content to the network. 

Serving a big client base requires enormous amounts of server resources. As demand grows, performance deteriorates and the system becomes fragile. Moreover, such an architecture is inherently fragile. Every single source of content at the servers is a potential single point of failure that can result in complete failure and lengthy downtime of the system. 

To tackle such flaws, technologies like Content Delivery Networks (CDNs)~\cite{cdn-survey} emerged to aggregate and multiplex server resources for many sources of content.
This way, a sudden burst of traffic could be more easily handled by sharing the load. Such innovations made the early client-server architecture a little more robust, but at considerable cost regarding infrastructure. Still, despite its inefficiency, the client-server model remains dominant today and runs most of the web. 

The  InterPlanetary File System (IPFS) \cite{ipfs-filecoin-2020} seeks to revert the historic trend of a client-server only Web.
It is a decentralized peer-to-peer content-addressed distributed file system 
that aims to connect all computers offering the same file system. Due to its decentralized nature, IPFS is intrinsically scalable. As more nodes 
join the network and content demand increases, so does the resource supply.
Such a system is inherently fault tolerant and, leveraging economies-of-scale, 
actually performs better as its size increases. It uses Merkle DAGs\cite{merkledag}\cite{merkledag2}, data structures to provide immutable, tamper-proof objects that are content addressed.

However, there are some crucial Content Delivery Network enabling features
are lacking in IPFS. In particular, the system lacks the 
capability of swiftly and organically approximate content 
from the request path, reducing the latency incurred by   future requests. 
It also does not prepare for the provider of an object to serve a sudden flood
of requests, thus rendering content actually inaccessible to some/many.

The goal of this work is, by taking advantage of the ecosystem 
build by IPFS, to develop an extension to IPFS that implements an adaptive
distributed cache that will improve the system's performance, further evolving it.
We aim to:

\begin{itemize}
  \item reduce the overall latency felt by each peer;
  \item increase the peers' throughput retrieving content;
  \item reduce the system's overall bandwidth usage;
  \item improve the overall balance in serving popular content by peers;
  \item finally, improve nodes' resilience to flash crowds.
\end{itemize}

In summary, Startrail serves as a key enabling component for an  IPFS-based CDN.


The rest of the document is organized as follows.  Section ~\ref{sec:arch}
describes the proposed solution. We address the architecture of IPFS
as a starting point to better understand the integration points of the proposed solution. Next, we further describe Startrail's architecture, its algorithms and data structures. Section~\ref{sec:eval} describes the testbed platform used, all the 
relevant metrics and obtained results. 
In Section~\ref{sec:rw}, we review previous work and systems in relevant related topics.
Some concluding remarks and extension proposals are presented in Section~\ref{sec:conc}.



\section{Architecture}
\label{sec:arch}

We designed Startrail to be an adaptive network cache, one that continually 
moves content ever closer to a growing source of requests. The goal is reducing 
on average the time it takes to access content on the network. 
It does so without requiring intermediate nodes to previously request such content, thus enabling smaller providers to serve bigger crowds, this way addressing the cost and infrastructure barriers-to-entry of typical CDNs.
It also aims to be interoperable, i.e. nodes running 
Startrail need not depend on other nodes to be effective, enabling  nodes
to contribute to the network even when adoption is not total.

\begin{figure}
  \centering
  \includegraphics[width=0.9995\linewidth]{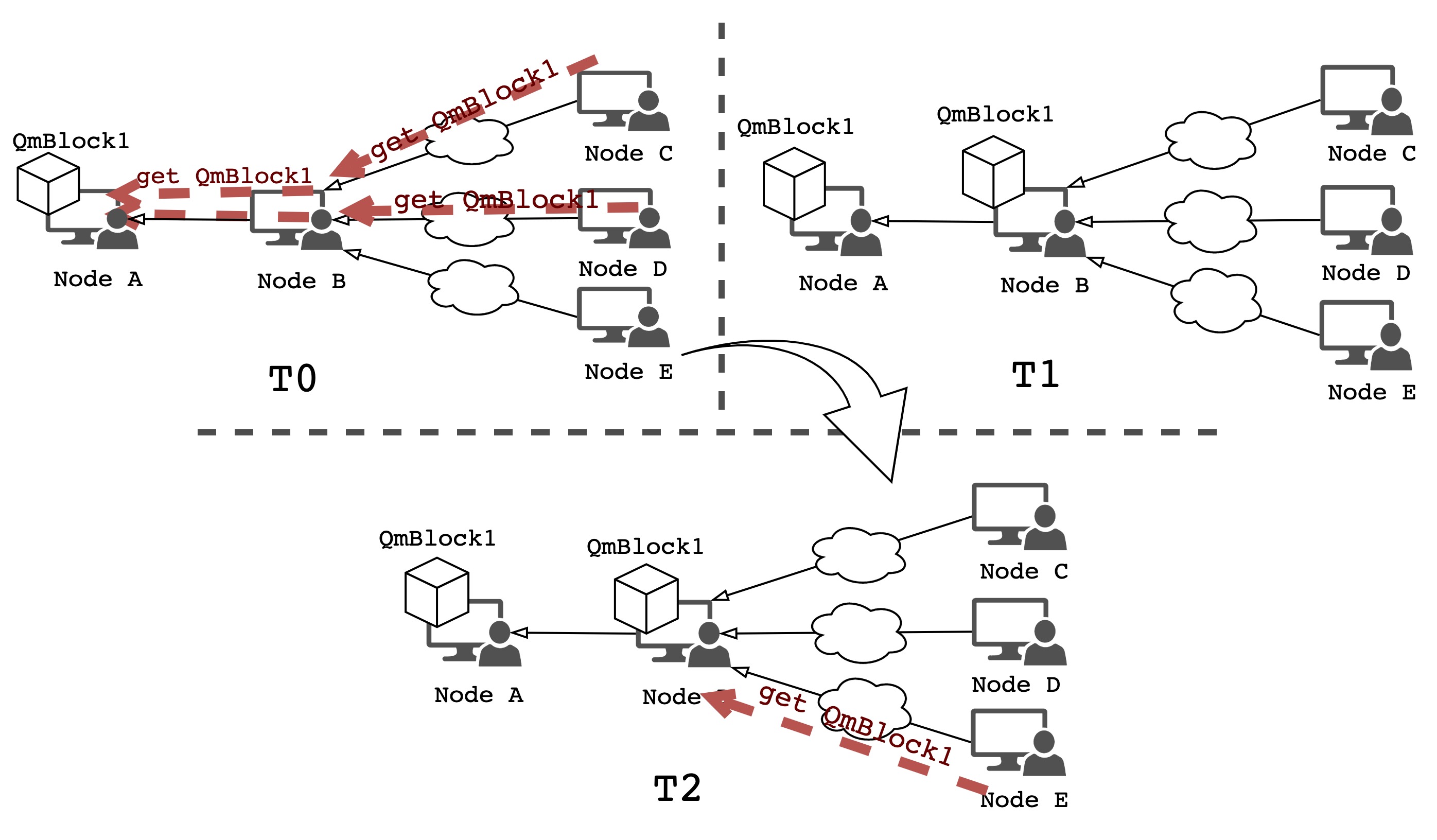}
  \caption{Startrail prototypical example flow}
  \label{fig:startrailFlowDraft}
\end{figure}

Figure~\ref{fig:startrailFlowDraft} depicts one of simplest scenarios with
a small portion of a network where all the nodes are running Startrail:
the content, in this case block \textit{QmBlock1}, is stored on \textit{Node A}.
Nodes \textit{C} and \textit{D} request \textit{QmBlock1} to the network.
While doing so, \textit{Node B} that is requested by both, detects that the
Content Identifier (CID) is popular and flags it, fetching and caching the content itself.
Later, when \textit{Node E} requests the block, the response needs not
traverse the whole network, it can  be served by \textit{Node B}.

\subsection{IPFS Core Architecture}

\begin{figure}
  \centering
  \includegraphics[width=0.75\linewidth]{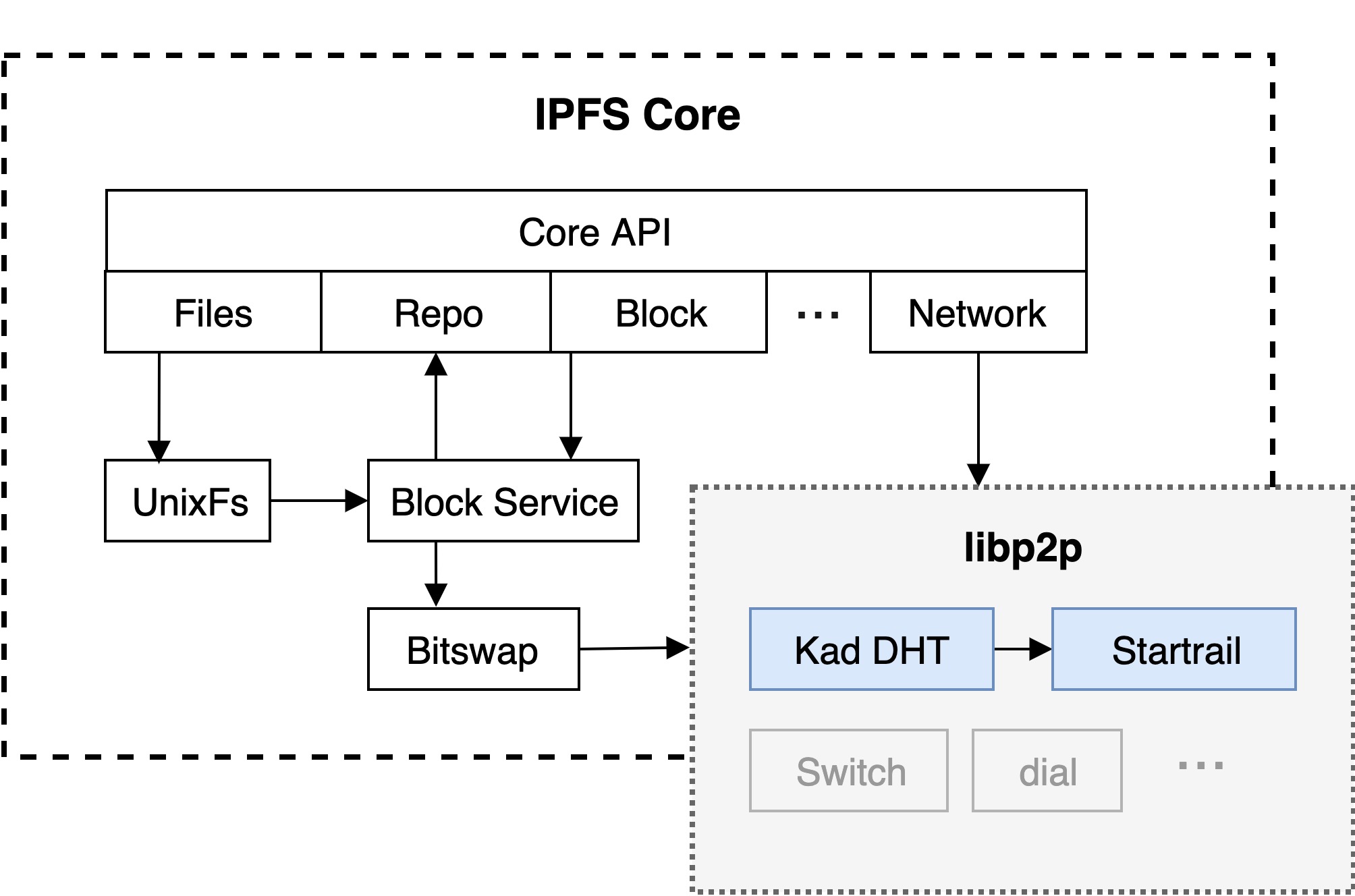}
  \caption{IPFS Core's Architecture}
  \label{fig:ipfsArchitecture}
\end{figure}

Startrail is built on top of IPFS and integrates with some of its deeper
internals and mechanics, thus we  address its core.
Objects on IPFS consist of Merkle DAGs of content-addressed
immutable objects with links, with a construction similar but more general than
a Merkle tree \cite{Merkle1988}. Deduplicated, these do not need to be balanced,
and non-leaf nodes may contain data. Since they are addressed by content, Merkle DAGs grant tamper proof, which is key to ensure the cached content is genuine.
A high-level overview of the architecture of the IPFS core is depicted in 
Figure~\ref{fig:ipfsArchitecture}. We shall delve into each of the illustrated 
components:

\begin{itemize}
  \item \textbf{Core API}: the Application Programing Interface (API) exposed    by the core, imported by both the Command Line Interface (CLI) and the HTTP   API;
  \item \textbf{Repo} - API responsible for abstracting the datastore 
    or database technology (e.g. Memory, Disk, AWS S3). It aims to enable
    datastore-agnostic development, allowing datastores to be swapped seamlessly;

  \item \textbf{Block}: API used to manipulate raw IPFS blocks;

  \item \textbf{Files}: API used for interacting with the File System.  textbf{UnixFS} is the engine for Unix files layout and chunking mechanisms for network file exchange;

  \item \textbf{Bitswap}: data trading module for IPFS. 
    It manages requesting and sending blocks to and from other peers in the network.
    Bitswap has two main jobs:
     i) to acquire blocks requested by the client from the network;
      and ii)  to  send blocks it holds to peers who want them;

  \item \textbf{BlockService}: content-addressable store for blocks, providing 
    an API for adding, deleting, and retrieving blocks.  This service is supported 
    by the \textbf{Repo} and \textbf{Bitswap} APIs.

  \item \textbf{libp2p}: networking stack and modularized library that grew out of IPFS. 
    It bundles a suite of tools to support the development of large scale peer-to-peer
    systems. It implements typical P2P mechanisms, e.g. \texttt{discovery}, \texttt{routing}, \texttt{transport}
    through many specifications, protocols and libraries. One  such library
    is \textbf{Kad-DHT}: the module responsible for implementing the
        Kademlia DHT~\cite{kademlia-02} with the modifications proposed by S/Kademlia \cite{Baumgart2007}.
        It has tools for peer discovery and content or peer routing.
\end{itemize}

Data exchanges on IPFS are handled by \textbf{Bitswap}~\cite{ipfs-bitswap-2021}, a \textbf{BitTorrent}
inspired protocol. Peers operate two data-structures: 

\begin{itemize}
  \item \textbf{\texttt{want\_list}}: set of blocks a node is looking to acquire;
  \item \textbf{\texttt{have\_list}}: set of blocks a node offers in exchange.\
\end{itemize}

\begin{figure}
  \centering
  \includegraphics[width=0.85\linewidth]{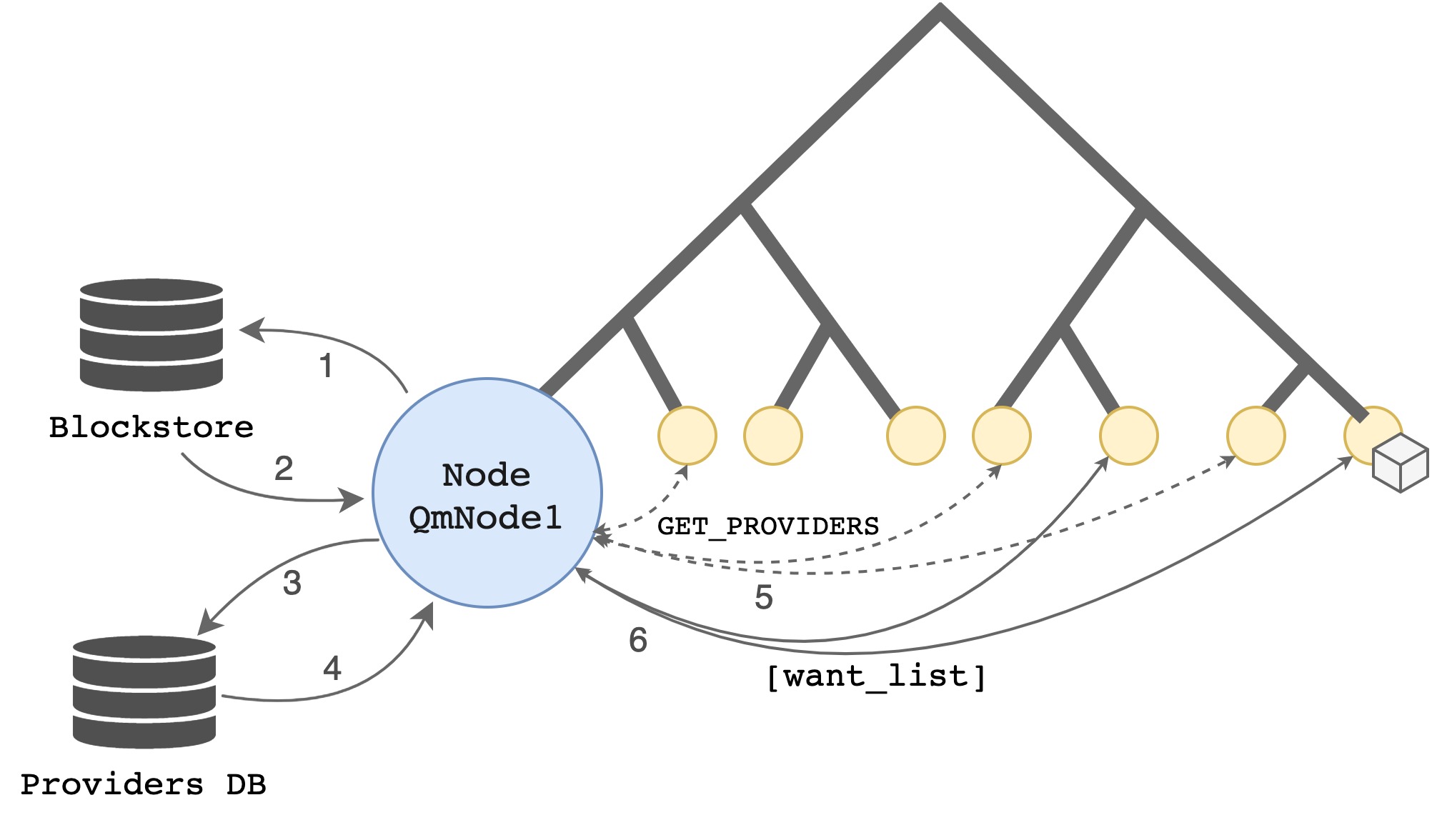}
  \caption{\textit{Bitswap}-\texttt{kad-dht} interaction when fetching a block from the network}
  \label{fig:ipfsGetBlockFlow}
\end{figure}

\vspace{3pt}
When searching for a block, \textbf{Bitswap} will first search the local \textbf{BlockService} for it. If not found, it will resort to the content routing module, in our case, the \textbf{Kad-DHT} module. This  will query the local providers database for the known providers of a certain CID and if not enough are gathered, query the rest from the DHT.
Once obtained the  set of potential providers, the node will
connect to them and pass  its \textbf{\texttt{want\_list}} containing the target CID (illustrated in Figure~\ref{fig:ipfsGetBlockFlow}).

\subsection{Startrail's Extension Architecture}

We now proceed to describe how
and where we integrate the Startrail cache as an IPFS extension to enable  CDNs over IPFS.
The first step is to identify where to tap into so that  new
content requests are intercepted. On IPFS,  one can do that via two different ways:

\begin{itemize}
  \item On \textbf{Kad-DHT}, checking for the CID associated to \texttt{GET\_PROVIDER} messages;
  \item On \textbf{Bitswap}, listening for new \texttt{want\_list} messages and tracking each of the included CIDs;
\end{itemize}

Our solution implements the former as it requires addressing just one type of messages. 

\begin{figure}
  \centering
  \includegraphics[width=0.85\linewidth]{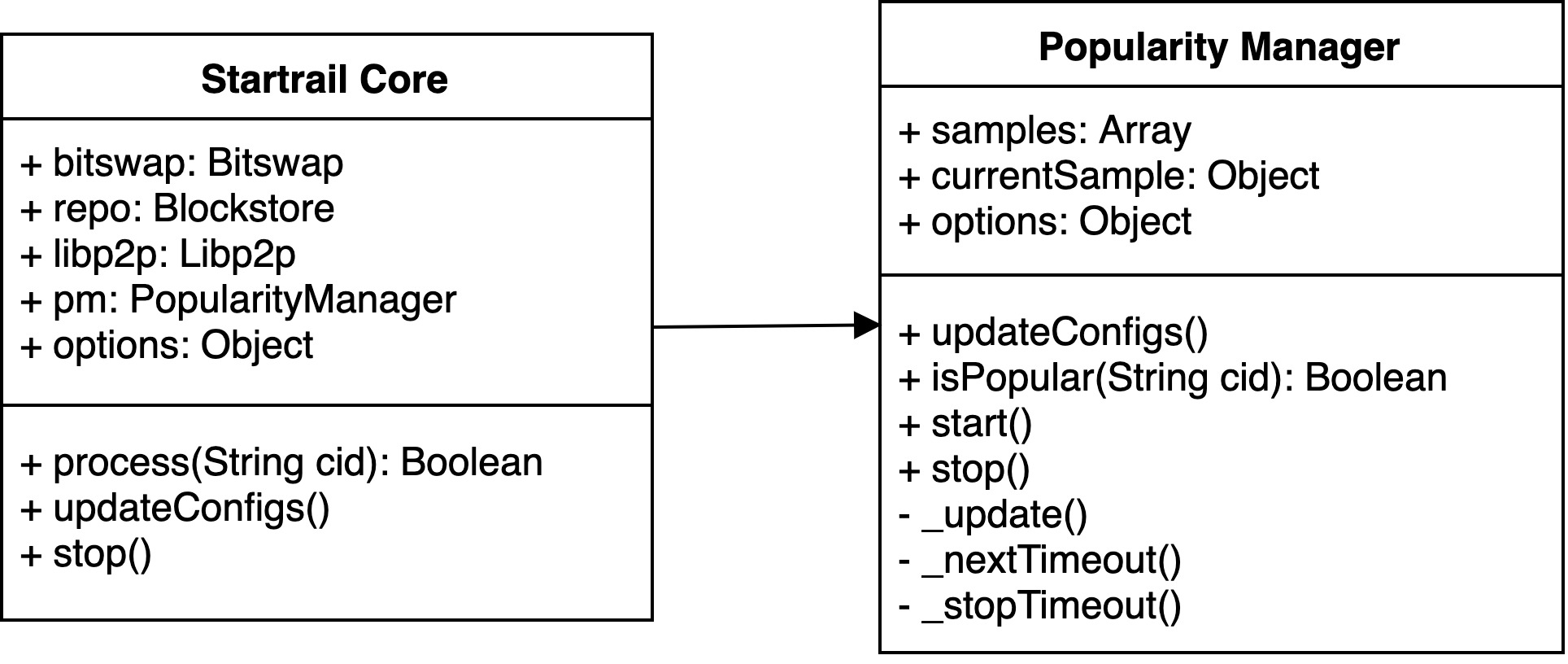}
  \caption{Class diagram of Startrail's Core and Popularity Manager}
  \label{fig:umlStartrail}
\end{figure}

Startrail's main activity is to detect trends in object accesses. To do so,
it uses two separate components, whose simplified class diagram is depicted in Figure~\ref{fig:umlStartrail}:

\begin{itemize}
  \item \textbf{Startrail Core}: exposing the Startrail API, it is 
     the interface other components will consume to work with StarTrail, It 
    is responsible for integrating with the data trading module (\textbf{Bitswap}), with
    the \textbf{BlockService} to access data storage, and libp2p for varios network utilities.
  \item \textbf{Popularity Manager}: component responsible for tracking objects' popularity. It is totally configurable and it can operate with any specified caching strategy.
\end{itemize}

The \textbf{Core} includes two main exposed functions: \texttt{process(cid)}, responsible for triggering the orchestration and popularity calculation, returning a Boolean for ease of integration with Kad-DHT; \texttt{updateConfigs()}, used for configuration refresh, fetching new configurations from the IPFS Repo and if changes are detected, will update the live ones (useful for hot realoading configurations when testing).

In the \textbf{Popularity Manager} class we find: \texttt{isPopular(cid)}, for updating and calculating the popularity for any CID passed as argument;
\texttt{updateConfigs()}, serving the same purpose as the above mentioned one;
\texttt{nextTimeout()}, managing the sampling timer responsible for scheduling timeouts; \texttt{update()}, runs every time the timeout pops, pushing the current sample to the sampling history and a new one is created.

\vspace{2pt}
\subsubsection{\textbf{Message processing algorithm}}

\begin{figure}
  \centering
  \includegraphics[width=0.95\linewidth]{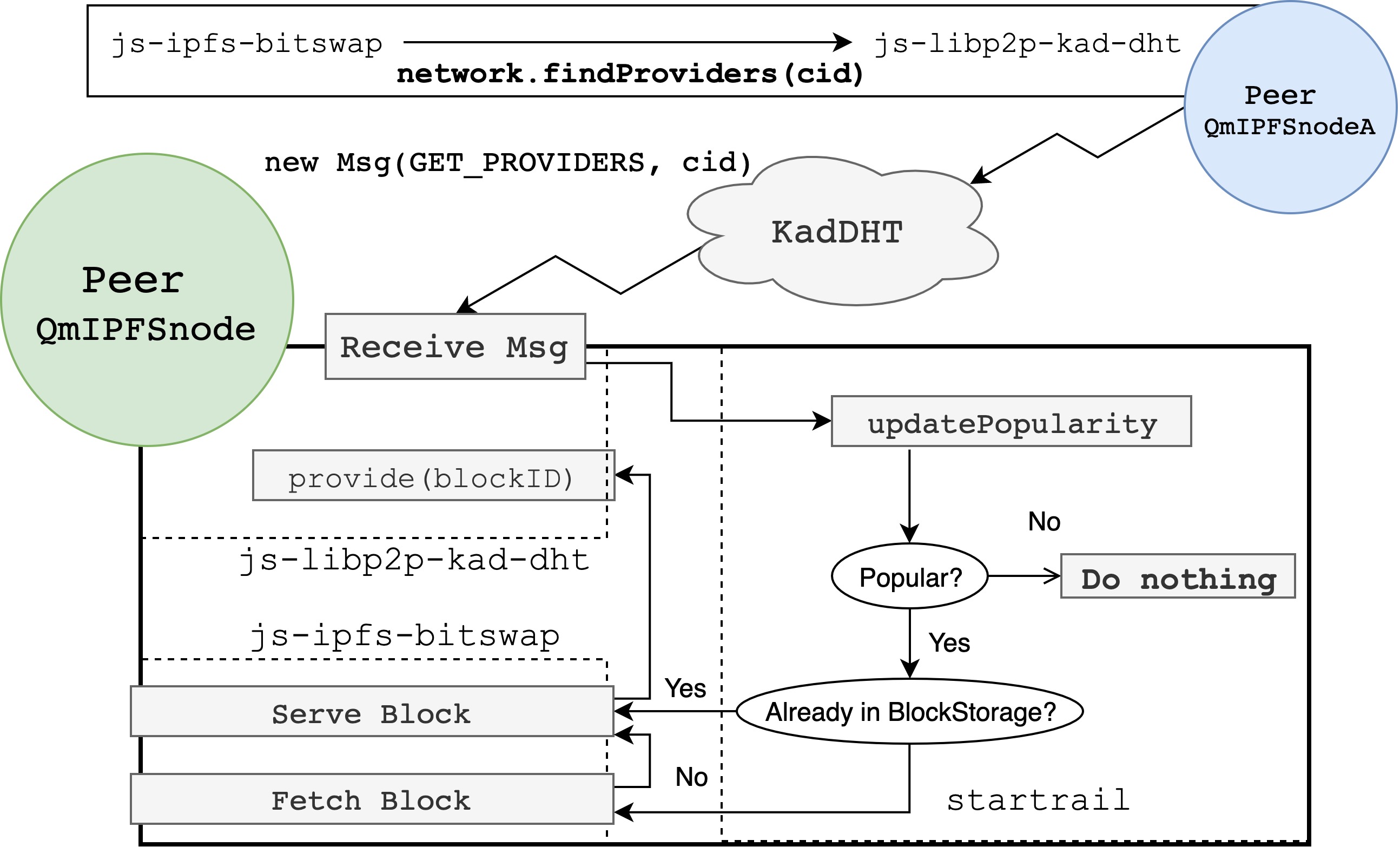}
  \caption{Execution flow inside the Startrail module}
  \label{fig:startrailFlow}
\end{figure}

To recognize patterns in object accesses, Startrail examines the CIDs sent on discovery messages. Hence, the \texttt{process()} function is triggered every time the \texttt{GET\_PROVIDER} message handler is called.
Figure~\ref{fig:startrailFlow} depicts the execution flow starting when a peer requests a block from the network triggering the search for providers on the network. Upon receiving a \texttt{GET\_PROVIDER} message, the {Kad-DHT} handler will execute Startrail's \texttt{process} hook. 

Following the popularity update, either no further action is required, or the block is flagged popular and the peer will attempt to fetch it or retrieve it from the BlockStorage. The block could potentially be found in the storage because, since we are using the IPFS BlockStorage, the block could have been previously fetched by either Startrail or the peer itself. Either way, subsequently to acquiring the block, the peer announces to the network that it is now providing it. 


\subsubsection{\textbf{Popularity Calculation Algorithm}}

By studying the current and past popularity of a certain CID,
we are able to likely forecast content that is going to, at least likely, remain popular in the future. Caching this locally and serving it to other peers has the benefit of making other nodes' accesses faster.
For simplicity, our forecast takes into consideration only a small subset of the 
node's \textit{past}. This subset, or window, can be obtained through various 
techniques. The one implemented in our solution is a hopping window. Here, sampling windows may overlap. This is desirable in our solution as we want to maintain some notion of continuity and smoothness between samples. Meaning that an object that was popular in the window before, still has high probability to remain popular in the current one, since a portion of the data remains the same. Although  parameters are fully configurable, the  defaults are 
{30 seconds} for
window duration, with hops of 
{10 seconds}.

\begin{figure}
  \centering
  \includegraphics[width=0.95\linewidth]{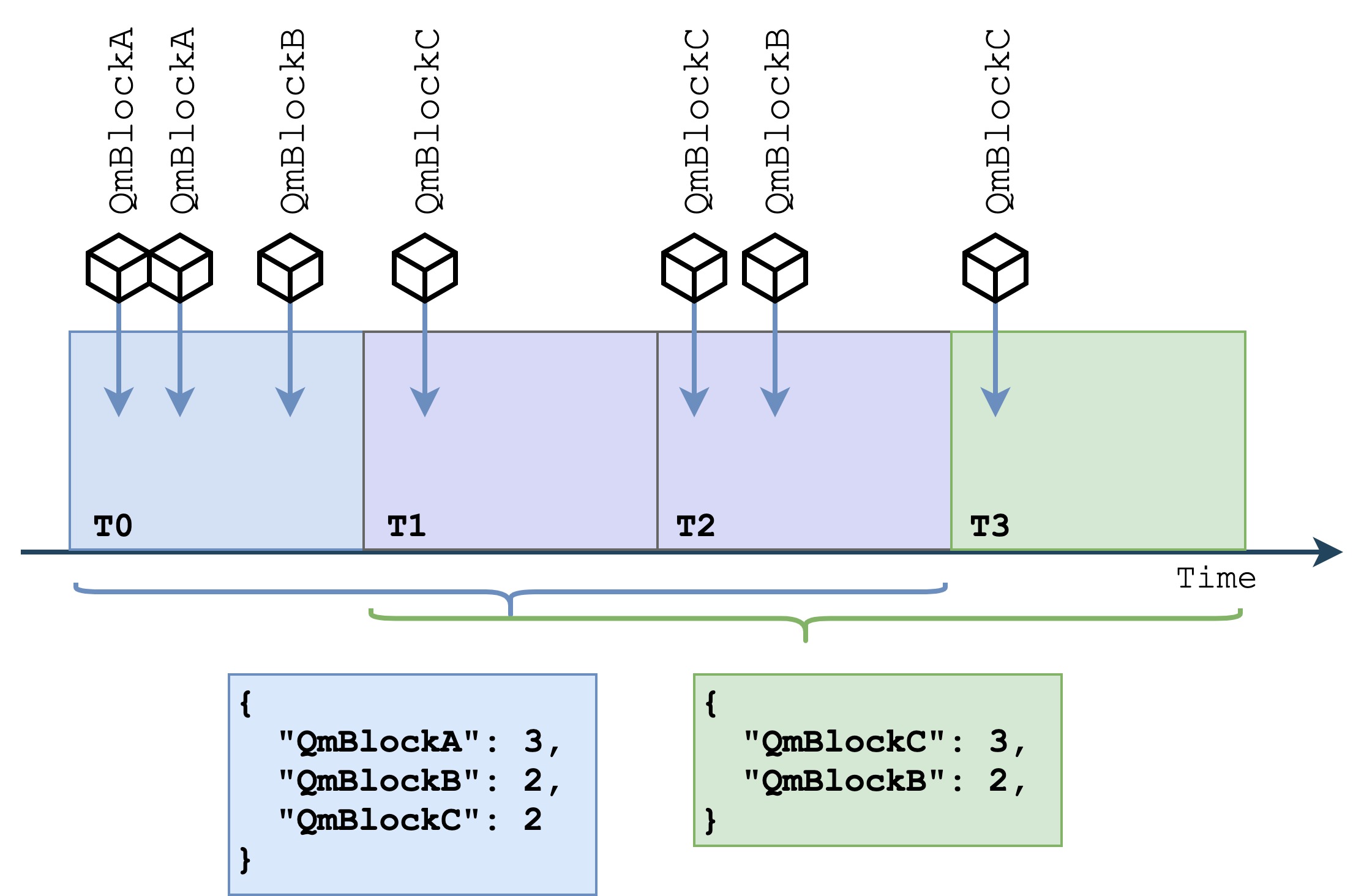}
  \caption{Interaction between new block arrivals and sampling windows}
  \label{fig:popularityManager}
\end{figure}

The Popularity Manager implements the hopping window by dividing it into hop-sized
samples. In our case we divide the total 30 second sampling  window into three 10 
second samples. 
Every time a new message is processed, the Startrail Core checks the popularity of
the referenced object by running the \texttt{isPopular()} function. The function will keep track
of objects it has seen in the current 10 second window; incrementing a counter
every time the CID processed. Every 10 seconds the current
window, or \texttt{sample} expires and is pushed onto a list that holds the previous
ones. It is on this latter list of samples (\texttt{samples} in the class diagram 
Fig.~\ref{fig:umlStartrail}) that the popularity calculations are made.
An illustration of the interaction between samples and block arrivals is represented
on Figure~\ref{fig:popularityManager}.

To calculate a block's popularity the Popularity Manager will first select the 
three most recent samples after concatenating the current one to this list. Next, it
will reduce the array outputting the total amount of times the object was observed.
If bigger than a certain configurable threshold the object is considered popular.
This heuristic has the benefits of (i) being fairly simple to implement and compute;
it also (ii) reacts quickly to changes in content access trends. In out current configuration, it can be considered
rather optimistic, since spotting the same object twice will consider it popular.

Our heuristic does not take into account the size of the content being cached. This
was a conscious decision, the reasoning for it is that on IPFS most blocks have the
maximum default size of 256Kb. Usually only the last one of the sequence that makes up a file
is smaller than that. Hence, we disregarded the block size as parameter for caching heuristic.

\subsubsection{\textbf{Cache Maintenance}}

In traditional caching systems \cite{Ali2011}, content is always cached by default with  the cache replacement algorithm responsible for discarding less relevant documents to create space for new ones. Contrarily, Startrail employs an heuristic to judge which objects should be cached in first place.

We deseign Startrail so that, for the most part, it works by leveraging the internal IPFS mechanics. This ensures the component is lightweight and uses the same procedures as the rest of the system. On Startrail, we allow the node to utilise the full amount of allocated storage by IPFS which defaults to 
{10 GB}. Once it  fills up, the IPFS garbage collector (GC)  discards unnecessary objects. The IPFS's GC removes the non-pinned objects. Hence, to prevent popular blocks from being collected when the it executes, we pin the popular objects. 

When a block stops being popular it is unpinned, leaving it at the mercy of the GC. When the threshold of 90\% of IPFS' storage is reached blocks are no longer pinned in order to leave room for new blocks.

\section{Evaluation}
\label{sec:eval}

\subsection{The Testbed}

There was a considerable amount of effort put into developing a testbed capable of simulating a realistic network. 
For our specific testbed we were looking for a solution that could fulfill the 
following requirements:

\begin{itemize}
  \item enable us to seamlessly adjust and change network conditions, e.g. latency, jitter;
  \item provide a platform to  gather and monitor diverse metrics
    and logs;
  \item scale  as more computing power is added to the testbed;
  \item effortlessly enable us to orchestrate and coordinate peers in the networks, i.e
    execute commands;
  \item allow for  integration with the IPFS codebase. This excluded possible alternatives    like PeerSim \cite{Jelasity2009}, as this would require porting the whole protocol 
    to the Java API.
\end{itemize}

\subsubsection{\textbf{Testbed Architecture}}
\label{sec:testbedArchitecture}

\begin{figure}
  \centering
    \includegraphics[width=0.95\linewidth]{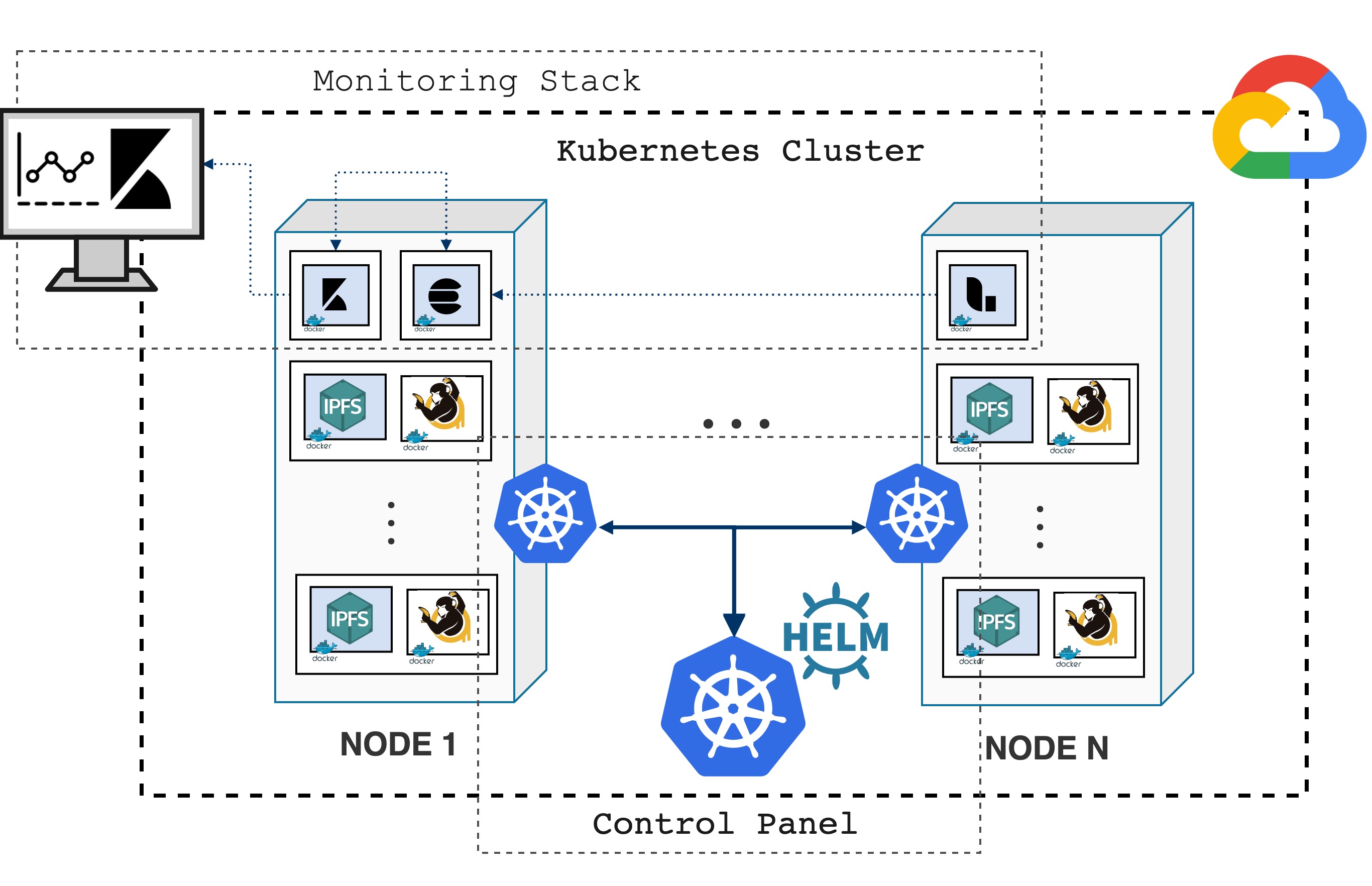}
    \caption{Architecture overview of the testbed network}
    \label{fig:testbedArchitecture}
  \end{figure}

The complete testbed architecture is depicted on Figure~\ref{fig:testbedArchitecture}.
To make the simulations easily reproducible,  we leveraged Helm\footnotemark, a tool that helps us release and manage Kubernetes applications. This enabled us to create different node configurations, \textit{Charts}, and change them for every deployment. \footnotetext{https://helm.sh}

To be able to gather data from different layers of the system during simulations, we resort to the ELK Stack\footnotemark~suite of tools. It
provides tools for log indexing, searching, transformation, storage and visualisation. \footnotetext{https://www.elastic.co/what-is/elk-stack}

\begin{figure}
\centering
  \includegraphics[width=0.75\linewidth]{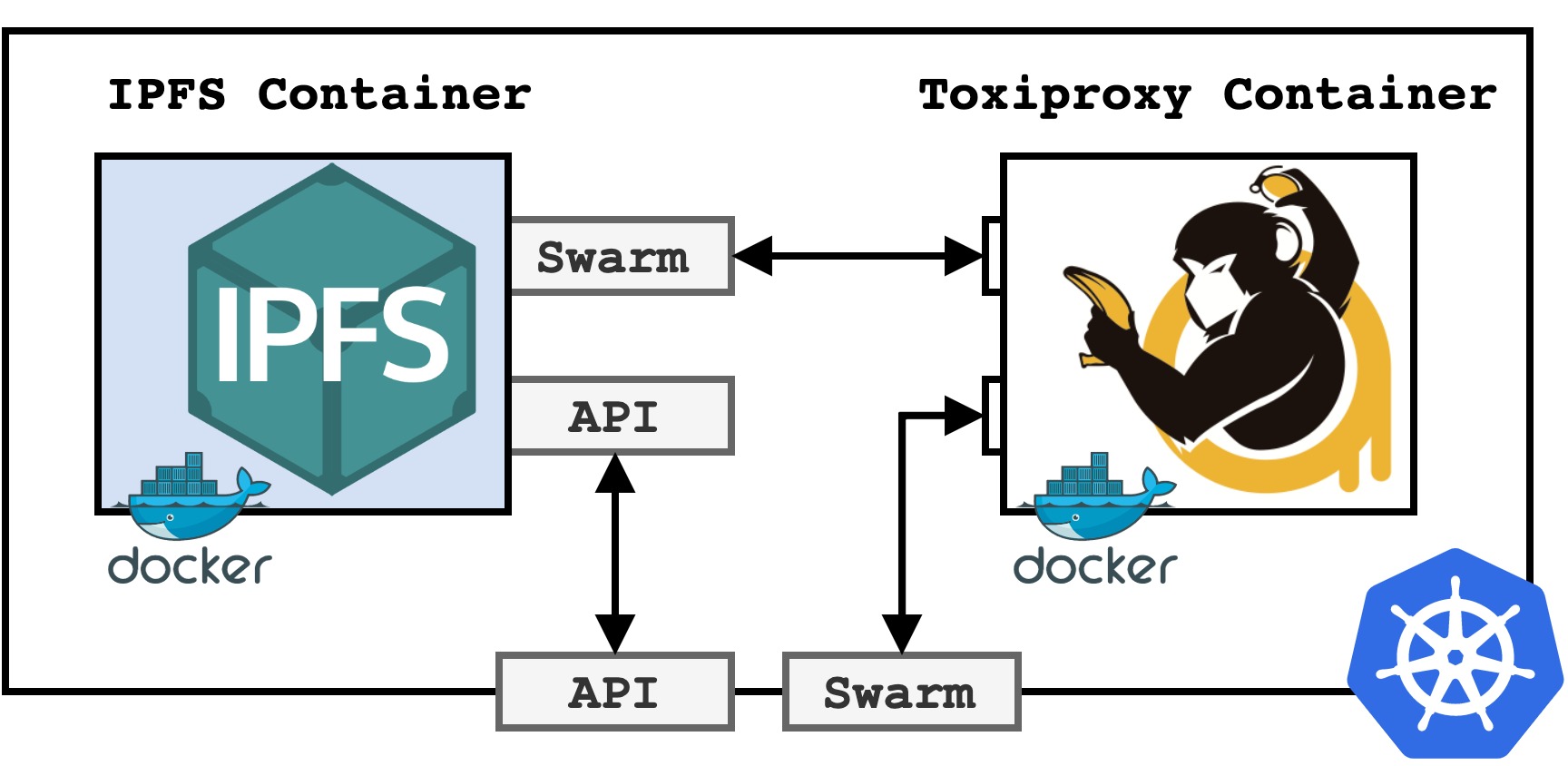}
  \caption{Composition of the Startrail Testbed Kubernetes Pod}
  \label{fig:kubernetesPod}
\end{figure}

To allow for easy addition  of  computing power into the testbed and 
management of deployments, we resorted to Kubernetes\cite{kubernetes}.
Kubernetes is a tool for deploying and managing containerized applications. 
It handles the discoverability and liveness of the deployed workloads. 

The smallest computing unit on Kubernetes is the \textit{Pod}. The \textit{Pod} is a formation of containers, in our case Docker \cite{docker} containers.
Our
\textit{Pod} architecture is illustrated in Figure~\ref{fig:kubernetesPod}.
Since  \textit{Pods} allows us to create any arbitrary composition of 
containers, we took advantage of this and injected, alongside every IPFS node, a \textit{Toxiproxy}\footnotemark~sidecar (a small proxy) from which we channel all IPFS traffic through, coming in and out from the node. This allows us to inject network variability into IPFS connections and simulate diverse network conditions. 
\footnotetext{https://github.com/Shopify/toxiproxy}

  \subsubsection{\textbf{Deploying a network}}

\begin{figure}
  \centering
    \includegraphics[width=0.95\linewidth]{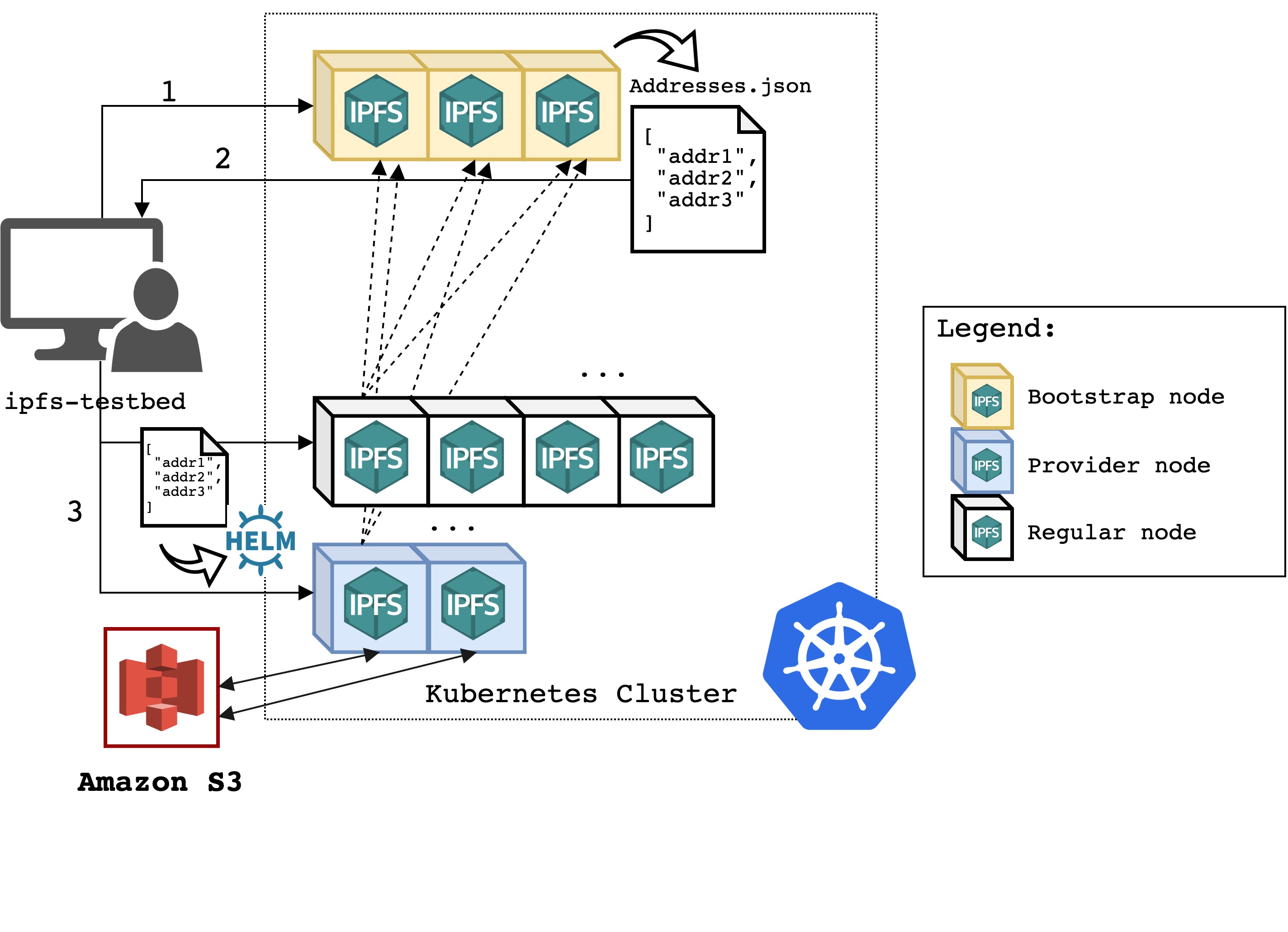}
    \caption{Deployment process of a new network on the Startrail testbed}
    \label{fig:ipfsTestbed}
  \end{figure}

  We resorted to Unix's Makefiles to automate the network setup as  much as possible, since we were aware that the network would have to be deployed many times. The deployment process, illustrated in Figure~\ref{fig:ipfsTestbed}, is as follows:
  \begin{itemize}
    \item setup Bootstrap nodes: on IPFS, to setup a network we first need to
      setup the bootstrap peers. These are used by other nodes as a \textit{Rendezvous Point}
      to join the network.
    \item create rest of nodes: additionally, using Helm's ability
      to dynamically configure releases, we need to direct these new nodes to the already
      setup Bootstrap ones.
    \item deploy Provider nodes: these are nodes preloaded with data. For these, as 
      datasets were sometimes of considerate dimensions, datasets were downloaded 
      onto the \textit{Pod} from an S3 Bucket before the starting the container.
  \end{itemize}
  
\subsubsection{\textbf{Interacting with the network}}

\begin{figure}
\centering
  \includegraphics[width=0.95\linewidth]{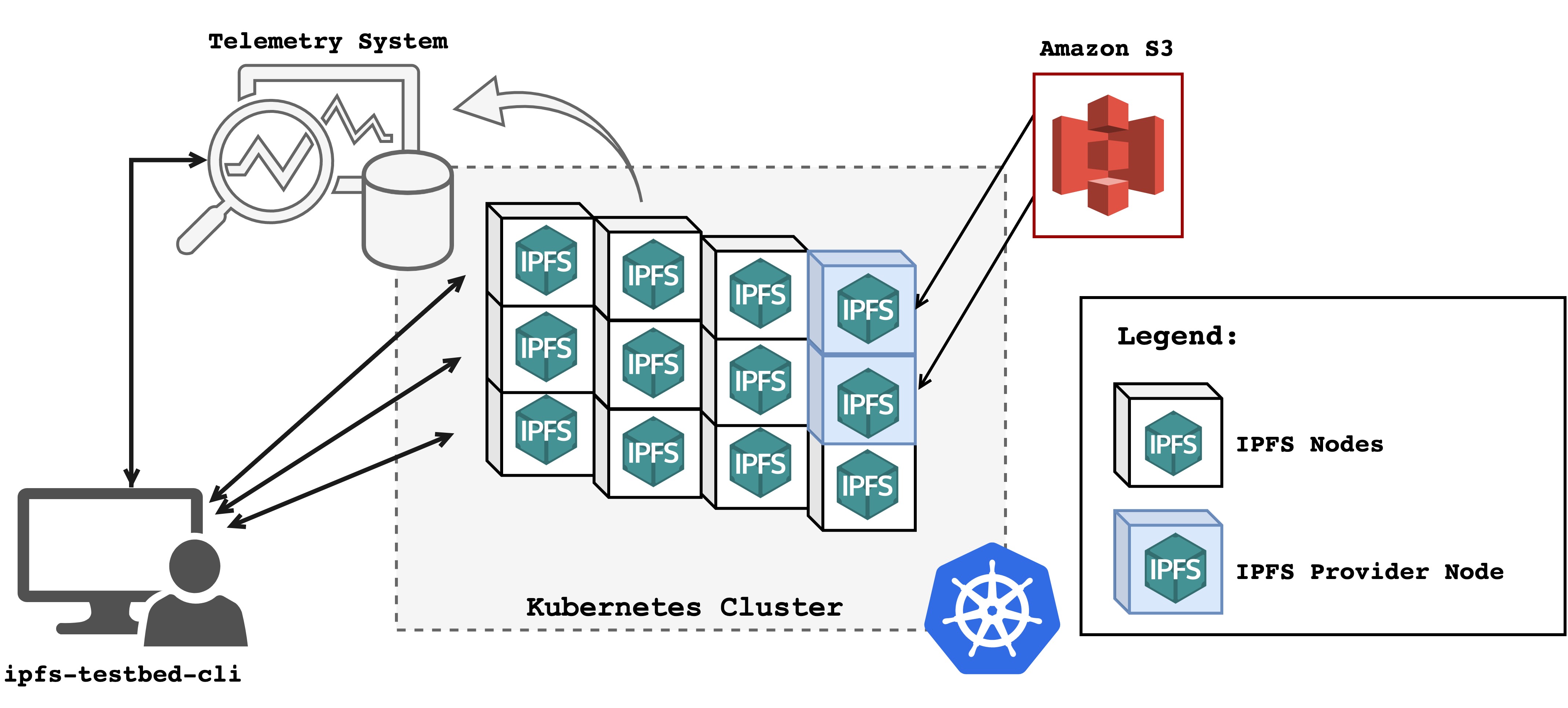}
  \caption{Testbed-cli orchestrating and monitoring Statrail testing cluster}
  \label{fig:cliFlow}
\end{figure}

To execute our tests, we developed a solution that utilizes the IPFS and \textit{Toxiproxy} APIs to orchestrate the peers and change the conditions of the network. This interaction can be examined in Figure~\ref{fig:cliFlow}.

With the network in place and a tool with which to execute tests on, it is
possible to execute multiple Startrail tests. We executed the tests from
our local machine that then would orchestrate commands to each individual peer according to the scripted test file. Using the monitoring platform we  then gather live data from the nodes.

\subsection{Relevant Metrics}
\label{sec:relevantMetrics}

The metrics  relevant for Startrail's  performance are:
\begin{itemize}
  \item \textbf{request duration}: the duration of network request is inherently
    the most important metric to analyze Startrail's impact on the system. It can
    assess if caching mechanism is working and how effective it is. Hence, to measure
    this we  analyze the 95th Percentile request duration.
  \item \textbf{memory usage}: considering that Startrail Nodes are caching content
    we want to analyze how much additional data each node has to store.
  \item \textbf{network usage}:  to assess how much each node has to resort
    to the network in order to fetch content. Hence, we measure the volume of traffic
     peers send to the network. 
\end{itemize}

For these metrics we calculate the 95th Percentile (95P), that is able to  exclude extreme values
from the average calculation, the  \textit{outliers}, as do typical cloud provider SLAs.

\subsection{Testing Setup}

The implementation network used to execute the simulations was deployed in the aforementioned Kubernetes cluster running on three \textit{n1-standard-4} Virtual Machines on Google Cloud Platform (GCP). These are a general-purpose type of machine
with 4 vCPUs, 15 GB of memory and 128 GB of storage, representative of many laptop, desktop, and smaller server machines powering IPFS. With this setup, we were able to simulate a network of 100 peers, of them 5  are bootstrap nodes, and 2 are providers preloaded with data.


To simulate different content access patterns we ran the following request management policies (implemented as functions in the test code) which pick a block from a dataset of thousands and order the peer to fetch it:
\begin{itemize}
  \item \textbf{Random Access (RA)}: the random access policy picks each block with equal
    probability. This pattern can serve as control;
  \item \textbf{Pareto Random (PR)}: with the Pareto Distribution \cite{Rosen1980}    serving as an adequate approximation for Internet objects popularity\cite{pareto}, this policy approximates  our desired distribution
    with an alpha equal to 0.3, meaning that 20\% of the blocks generate 80\%
    of the overall network traffic;
  \item \textbf{File Random (FR)}: here we divided the total list of blocks into smaller lists, each amounting to 3Mb in block data. Selecting one of these lists means that the peer  fetches all the nodes in the list, thus simulating complete file access. We  also use a Pareto distribution  here.
\end{itemize}

Table~\ref{Tab:testStartrailConditions} summarizes all the test scenarios simulated. Each one had an induced latency of 100 ms, ran for 10 minutes and each node requested a new block every 30 seconds.

\begin{table}
\vspace{5pt}
  \begin{center}
      \begin{tabular}{lclllllc}
        \hline
        \multicolumn{1}{c}{\textbf{Test Name}} & \textbf{Startrail} & \textbf{Access Type}  & \textbf{Window Size} & \textbf{Threshold} \\ \hline
        RA w/o Startrail                  & False              & RA                             & N.A.              & N.A.                  \\
        RA w/ Startrail                    & True               & RA                             & 3*10sec              & 2                  \\
        PR w/o Startrail                   & False              & PR              & N.A.              & N.A.                  \\
        PR  w/ Startrail                  & True               & PR              & 3*10sec              & 2                  \\
        FR w/o Startrail             & False              & FR                        & N.A.              & N.A.                  \\
        FR w/ Startrail               & True               & FR                        & 3*10sec              & 2                  \\ \hline
      \end{tabular}
    \caption{Different testing condition for running network tests}
    \label{Tab:testStartrailConditions}
  \end{center}
\end{table}
The parameter columns are:
\begin{itemize}
  \item \textbf{Startrail}: defines if Startrail was running on all the nodes 
    in the network; 
  \item \textbf{Access Type}: indicates which of the previously mentioned access 
    patterns we are simulating;
  \item \textbf{Latency}:  the amount of latency introduced by \textit{Toxiproxy};
  \item \textbf{Req. Freq.}: specifies at which  frequency the requests for blocks
    - or array of blocks, in the case of File Random - are made by the individual nodes;
  \item \textbf{Duration}: indicates for how long we run the simulation;
  \item \textbf{Window} and \textbf{Threshold}: specify the used Startrail configuration,
    if applicable. \textit{Window} being the amount of samples and the duration of
    each, in seconds; and the \textit{Threshold}  of accesses to trigger caching   in   Startrail.
\end{itemize}

\subsection{Tests Results}

\subsubsection{\textbf{Latency Analysis}}

\begin{figure}
  \centering
  \includegraphics[width=0.75\linewidth, height=6.5cm]{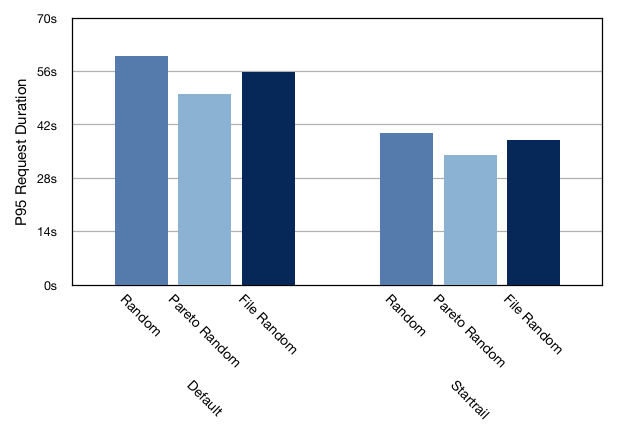}
  \caption{95th percentile of request duration for the different testing scenarios}
  \label{fig:startrailTests}
\end{figure}

The chart on Figure~\ref{fig:startrailTests} depicts the calculated 95P Request
duration for each simulation. The simulation of the random access running on a
regular IPFS nodes' network, on the far left, yielded a P95 latency of 60 seconds. The same simulation running on the Startrail network only took 40 seconds. This is a considerable reduction of one third. Similar gains in speed can be observed for the tests with Pareto distribution access pattern. 

At first glance, one would think that this would be the policy where the impact of Startrail would be the most evident, because the blocks would reach the cache threshold easier, In this case, however, since the same blocks are being requested more often, different peers on the network also serve the content because they previously downloaded it. Hence, the difference remains the same. For the file access type the proportion of gains remains similar, with the overall latency going up since now we are requesting a lot more of different blocks.

\subsubsection{\textbf{Memory Consumption Analysis}}

\begin{figure}
  \centering
  \includegraphics[width=0.75\linewidth, height=6.5cm]{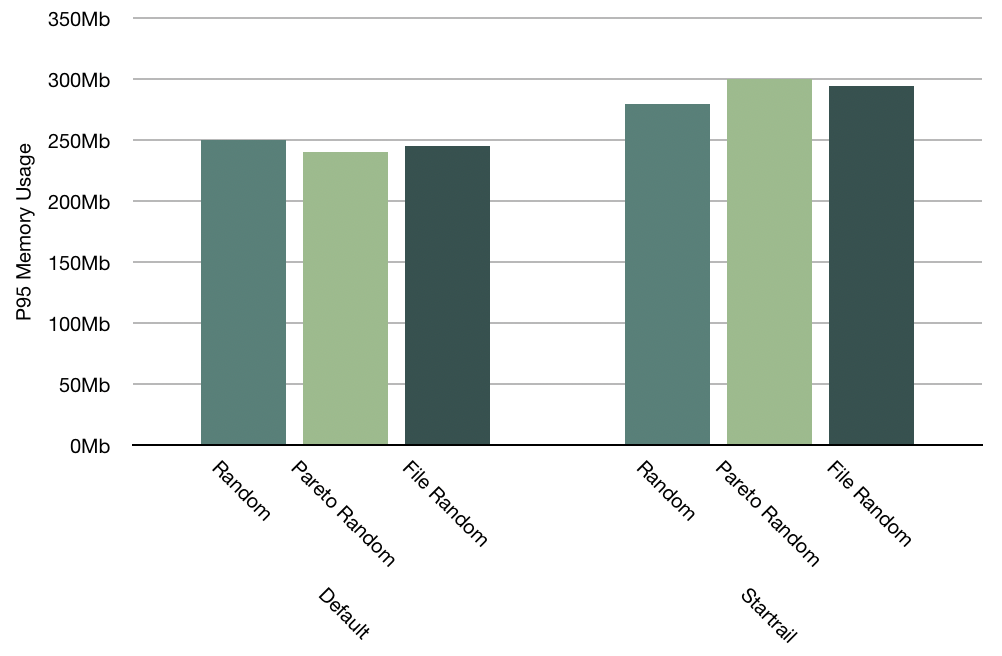}
  \caption{95th Percentile of memory usage on the different testing scenarios}
  \label{fig:startrailTestsMem}
\end{figure}

The chart on Figure~\ref{fig:startrailTestsMem} compares the memory cost of
running the simulations on a network with and without Startrail. Its analysis  reveals that running the simulations without Startrail costs generally the
same, with only slight variations proportional to the amount of different blocks
requested. 
For the simulations run on the Startrail network we observe a consistent increase in memory 
utilization. This is expected since nodes are now storing more content, 
the cached blocks.

For the Pareto access pattern we notice an increase in memory consumption relative to the all random ones. This is not completely intuitive, as the smaller diversity of blocks requested could mean less content being cached. However, we find  that in this simulation a smaller subset of the dataset is now being constantly requested, meaning that, although smaller, it reaches the cache threshold with a very high probability and is therefore all stored.
Additionally, because it doesn't grow significantly and past the IPFS default 10 GB of storage,
the GC is never triggered and thus the cached blocks are kept for the
whole duration of the test. This does not happen with the random access pattern
because, since it follows a uniform distribution, requests for the same block 
can be  scattered over time and thus some never reach the threshold.

\subsubsection{\textbf{Network Consumption Analysis}}

\begin{figure}
  \centering
  \includegraphics[width=0.75\linewidth, height=6.5cm]{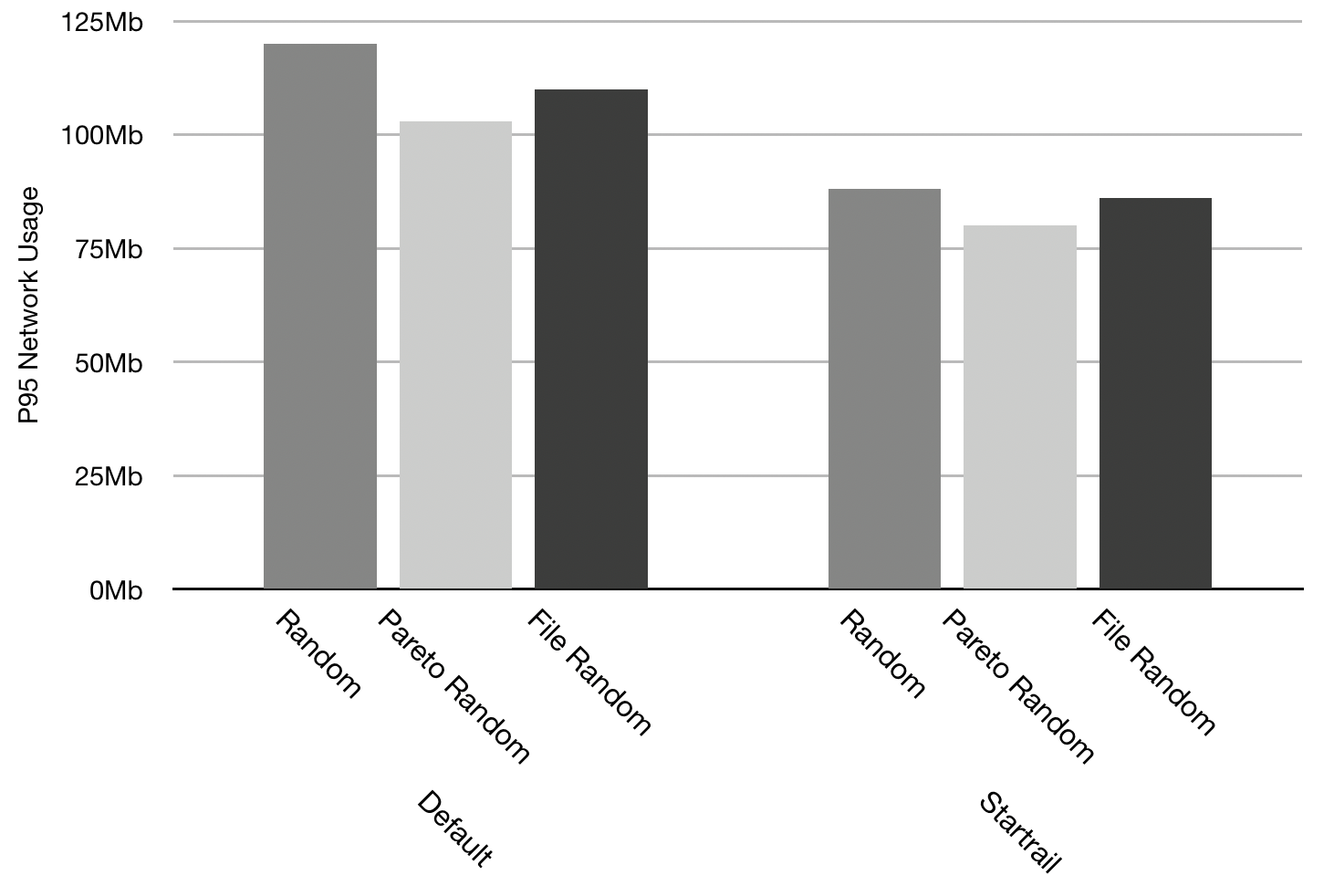}
  \caption{95th Percentile of network usage on the different testing scenarios}
  \label{fig:startrailTestsNetwork}
\end{figure}

The chart on Figure~\ref{fig:startrailTestsNetwork} illustrates the network impact of running simulations with and without Startrail.  Comparison of the obtained results reveals that the savings in network traffic (MBs) are proportional to the speed ups in request latency. This happens because requests are being served closer to their source by caching nodes and  thus fewer messages have to transit over the network.

\subsubsection{\textbf{Variable Startrail adoption}}

Additionally, we also analyzed the impact of the percentage of Startrail-enabled nodes in the network. This allows us to assess the interoperability of Startrail with the base IPFS and to confirm their compatibility, which is key for incremental deployment and adoption. Furtermore, it also allows us to determine whether  the benefits of Startrail are significant,  even if only a smaller percentage of nodes are contributing to caching and if, on the other hand, there is an optimal percentage of Startrail participation in IPFS that would yield the best  performance.

In order to evaluate which of the above mentioned propositions were true we ran aditional simulations. The conditions of these were similar to the ones presented earlier: in each test the latency induced was 100, the tests ran for 10 minutes and each node requested a new block every 30 seconds.
The distinction between the ones described before and these is that in the latter we alternated the percentage of nodes running Startrail that were deployed on the network. The percentages were: 0\%, 30\%, 50\%, 80\% and finally 100\% Startrail nodes.

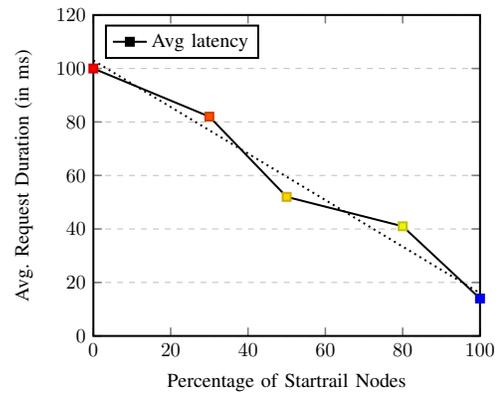
\begin{figure}
  \centering
  \begin{tikzpicture}[scale=.75]
    \begin{axis}[
        xlabel={Percentage of Startrail Nodes},
        ylabel={Avg. Request Duration (in ms)},
        xmin=0, xmax=100,
        ymin=0, ymax=120,
        xtick={0,20,40,60,80,100},
        ytick={0,20,40,60,80,100,120},
        legend pos=north west,
        ymajorgrids=true,
        grid style=dashed,
        line width=1pt,
        tick style={line width=0.8pt}
      ]

      \addplot[ scatter, mark=square* ]
      coordinates {
        (0,100)(30,82)(50,52)(80,41)(100,14)
      };
      \legend{Avg latency}
      \addplot[mark=none, black, scale=1, dotted]
      coordinates {
        (0,103)(100,16)
      };
    \end{axis}
  \end{tikzpicture}
  \caption{Avg request duration vs. Startrail nodes percentage}
  \label{fig:startrailTestPercentage}
\end{figure}

The results obtained from running the simulations were compiled into the chart  in
Figure~\ref{fig:startrailTestPercentage}. The  samples start on the far left
with higher values of request latency for no Startrail participation on the network,
and decrease nearly linearly as the percentage of Startrail nodes increases. 
Therefore, it appears that the impact of Startrail nodes' percentage on the network performance can be approximated 
through the linear regression drawn on the dotted line in the graph. This supports
our initial proposition that th performance would improve as the percentage of Startrail
nodes increases. 

Nevertheless there is a slightly higher slope in the 30\% to 50\%
range. That however does not allow any conclusions of an optimal point to be
taken (possibly due to some experimental noise or from the  coarse-grained range of the parameter used in the testing but that still highlights the relevant trend).

\section{Related Work}
\label{sec:rw}

Relevant work related to Startrail includes distributed storage and file  systems (in particular peer-to-peer), content delivery networks, and web distributed technologies.


\paragraph*{\textbf{Distributed File and Storage Systems}}
A file system  supports  sharing of files over a set of network connected nodes. The challenge is keeping the system performant, secure and robust. In addition, file location and availability assume significance. One way of increasing the availability is by using data replication and in order to increase performance caching can also be used.
Distributed follow different types of architectures \cite{thanh2008b}.

The 
  \textit{Client-Server Architecture}, offers a directory tree to multiple clients (e.g.~NFSv3~\cite{Goldberg1985}). The capacity of the system is limited by the capacity of the server, it being  a single point of failure.
  In \textit{Object-based File Systems}, metadata and data management are kept separate. A master server maintains a directory tree and keeps track of replica servers. It only handles data placements and load balancing, with  data transfer carried out directly among nodes. This architecture allows for incremental scaling, e.g. GFS~\cite{gfs}, MDFS~\cite{MSDFS}.
  
  \textit{Peer-to-Peer storage systems} are distributed systems consisting of interconnected nodes (peers) serving content blocks. These can be: i)
   \textit{unstructured}, when the network imposes no constraints on the links between different nodes. The placement of content is completely unrelated to the overlay topology, requiring little maintenance while peers enter and leave the system. These systems lack a way to index data (thus, resource lookup mechanisms consist of inefficient brute-force or gossiping), and limited scalability; or ii) 
   \textit{structured}, that fix the scalability issues of unstructured. Here, the overlay topology is tightly controlled and data is placed at specific logical locations in the overlay \cite{El-ansary2004a}. These systems provide a mapping between the data identifier and location, so that queries can be efficiently routed to the node(s) with the desired data, e.g. Chord \cite{Stoica2001}, Pastry \cite{rowstron2001pastry}, Tapestry \cite{zhao2001tapestry}, and Kademlia~\cite{kademlia-02} that inspires the IPFS overlay. 

\paragraph*{\textbf{Content Delivery Networks}}
Content Delivery Networks (CDNs) are networks of geo-distributed machines that deliver web content to users based on their geographic location~\cite{cdn-survey}, including based on content naming~\cite{icn-disc-2017} and leveraging wide area network topology~\cite{icdn-ndn-2020}. Still, they  improve content delivery speed and service availability only at the cost of owning or renting the geo-distributed replicas which are mostly fixed.


\paragraph*{\textbf{Web Distributed Technologies}}  
A few technologies are key for the successful  design of decentralized systems in the Web. There are:
 i) the \textit{Browser}, a very powerful tool to fuel adoption. Building a solution that runs on the browser means one is able to reach a broad audience; ii) 
  \textit{Node.js} allows running JavaScript in servers. It bundles with NPM (Node Package Manager), a very power library of reusable JavaScript modules;
  iii)
   \textit{WebRTC} provides browsers and mobile applications with peer-to-peer Real-Time Communications capabilities, accessible through a Javascript API (with a suite of protocols addressing NAT traversal problems).

In summary, IPFS offers a file system API over a peer-to-peer block storage system and is fully powered by web technologies. However, it lacks the efficient content dissemination of CDNs.  Startrail enables it without additional cost of ownership or renting of traditional CDNS, while  being fully decentralized, adaptive and flexible.

\section{Conclusions}
\label{sec:conc}


In this paper  we presented  a solution that extends the existing IPFS and improves
it as a content sharing system and its ability to distribute it.
After identifying the key architectural elements and functionality of IPFS and its shortcomings, we developed  Startrail as an extension in the form of an adaptive caching component. 








The obtained results show that a
network running Startrail nodes is able to perform better than one running 
only IPFS nodes. Startrail reduces the request latency by 30\%, at the cost of small
increase in total memory consumption of 20\% while also reducing bandwidth 
utilization by around 25\%.

Additionally, we assessed the impact that different percentages of Startrail nodes
have in the overall network performance. The results, confirm the expectation that 
there is an inverse relation between Startrail nodes percentage and network latency. 
When one increases, then other is reduced.

\paragraph*{Future work}

Although the results are positive and bring improvements to IPFS's operation,
there are potential further advances to be
implemented. Startrail enables the development of additional enhancement features, e.g.:
%
i) Startrail takes advantage of a single
  request popularity probe, the discovery messages. In order to achieve a broader
  probing potential an improvement could be made allowing the system to further 
  analyze requests on the received \textit{want\_list}s;
ii) Caching thresholds on
  Startrail are static which can lead to caching imbalances when under high
  stress. One very interesting research topic would develop an heuristic that 
  would dynamically adapt to the amount of requests the node processes;
iii)  Prefetching 
  on IPFS can be implemented at system level or at application level. Startrail 
  is a key crucial enabler for its operation. Startrail makes it possible for 
  the prefetching mechanism to have a reduced impact on the network by leveraging 
  the caching, which, instead of stressing the network, has the effect of 
  setting up the network caches for traffic to come.

\vspace{6pt}\footnotesize{\textbf{Acknowledgements}
This work was supported by national funds through FCT, Fundação para a Ciência e a Tecnologia, under project UIDB/50021/2020.}

\bibliographystyle{unsrt}
\scriptsize{
\bibliography{ref.bib}
}

\end{document}